\documentclass[prb,superscriptaddress,twocolumn,showpacs]{revtex4}
\usepackage{amsmath}
\usepackage{amssymb}
\usepackage{bm}
\usepackage{epsfig}
\usepackage{graphicx}
\usepackage{color}
\usepackage{hyperref}
\usepackage[T2A]{fontenc}
\usepackage[english]{babel}

\begin{document}
\title{Exciton fine structure and spin decoherence in monolayers of transition metal dichalcogenides}
\author{M.M. Glazov}
\email[]{glazov@coherent.ioffe.ru}
\affiliation{Ioffe Physical-Technical Institute of the RAS, 194021 St. Petersburg, Russia}
\author{T. Amand}
\author{X. Marie}
\author{D. Lagarde}
\author{L. Bouet}
\author{B. Urbaszek}
\affiliation{Universit\'e de Toulouse, INSA-CNRS-UPS, LPCNO, 135 Av. de Rangueil, 31077 Toulouse, France}

%

\begin{abstract}
We study the neutral exciton energy spectrum fine structure and its spin dephasing in transition metal dichalcogenides such as MoS$_2$. The interaction of the mechanical exciton with its macroscopic longitudinal electric field is taken into account. The splitting between the longitudinal and transverse excitons is calculated by means of the both electrodynamical approach and $\bm k \cdot \bm p$ perturbation theory. 
This long-range exciton exchange interaction can induce valley polarization decay. The estimated exciton spin dephasing time is in the picosecond range, in agreement with available experimental data.
\end{abstract}


\maketitle


\textit{Introduction}. Monolayers (MLs) of transition metal dichalcogenides, in particular, MoS$_2$ form a class of novel two-dimensional materials with interesting electronic and optical properties.
The direct band gap in these systems is realized at the edges of the Brillouin zone at points $\bm K_+$ and $\bm K_-$.\cite{Xiao:2012cr} Strikingly, each of the valleys can be excited by the radiation of given helicity only.\cite{Cao:2012a,Mak:2012a,Zeng:2012ys} Recent experiments have indeed revealed substantial optical orientation in ML MoS$_2$ related to selective excitation of the valleys by circularly polarized light.\cite{Sallen:2012qf,Kioseoglou,PhysRevLett.112.047401,Zhu:2013ve} Strong spin-orbit coupling in this material lifts the spin degeneracy of electron and hole states even at $\bm K_+$ and $\bm K_-$ points of the Brillouin zone resulting in relatively slow spin relaxation of individual charge carriers, which requires their intervalley transfer.\cite{Li:2013qf,Molina-Sanchez:2013mi,Ochoa:2013wd,MWWuMoS2,Song:2013uq} 
However recent time-resolved measurements revealed surprisingly short, in the picosecond range, transfer times between valleys.\cite{doi:10.1021/nl403742j,Wang:2013hb} This could be due to excitonic effects which are strong in transition metal dichalcogenides\cite{Cheiwchanchamnangij:2012pi,Mak:2013lh}
Although individual carrier spin flips are energetically forbidden, spin relaxation of electron-hole pairs can be fast enough owing to the exchange interaction between an electron and a hole forming an exciton,\cite{Yu:2014uq,Yu:2014fk} in close analogy to exciton dephasing in quantum wells.\cite{maialle93}

Here we study the energy spectrum fine structure of bright excitons in MoS$_2$ and similar compounds. We use both the electrodynamical approach where the interaction of the mechanical exciton (electron-hole pair bound by the Coulomb interaction) with the macroscopic longitudinal electric field is taken into account\cite{birpikus_eng,denisovmakarov} and $\bm k \cdot \bm p$ perturbation theory. We obtain the radiative lifetime of excitons as well as the splitting between the longitudinal and transverse modes. We compare the developed theory
based on long-range exchange interaction (in contrast to calculations in Ref.~\onlinecite{Yu:2014uq} based on short range interaction) with experimental results on spin decoherence in MoS$_2$ MLs\cite{PhysRevLett.112.047401,doi:10.1021/nl403742j,Wang:2013hb} and demonstrate an order of magnitude agreement between the experiment and the theory. 

\begin{figure}[t]
\includegraphics[width=\linewidth]{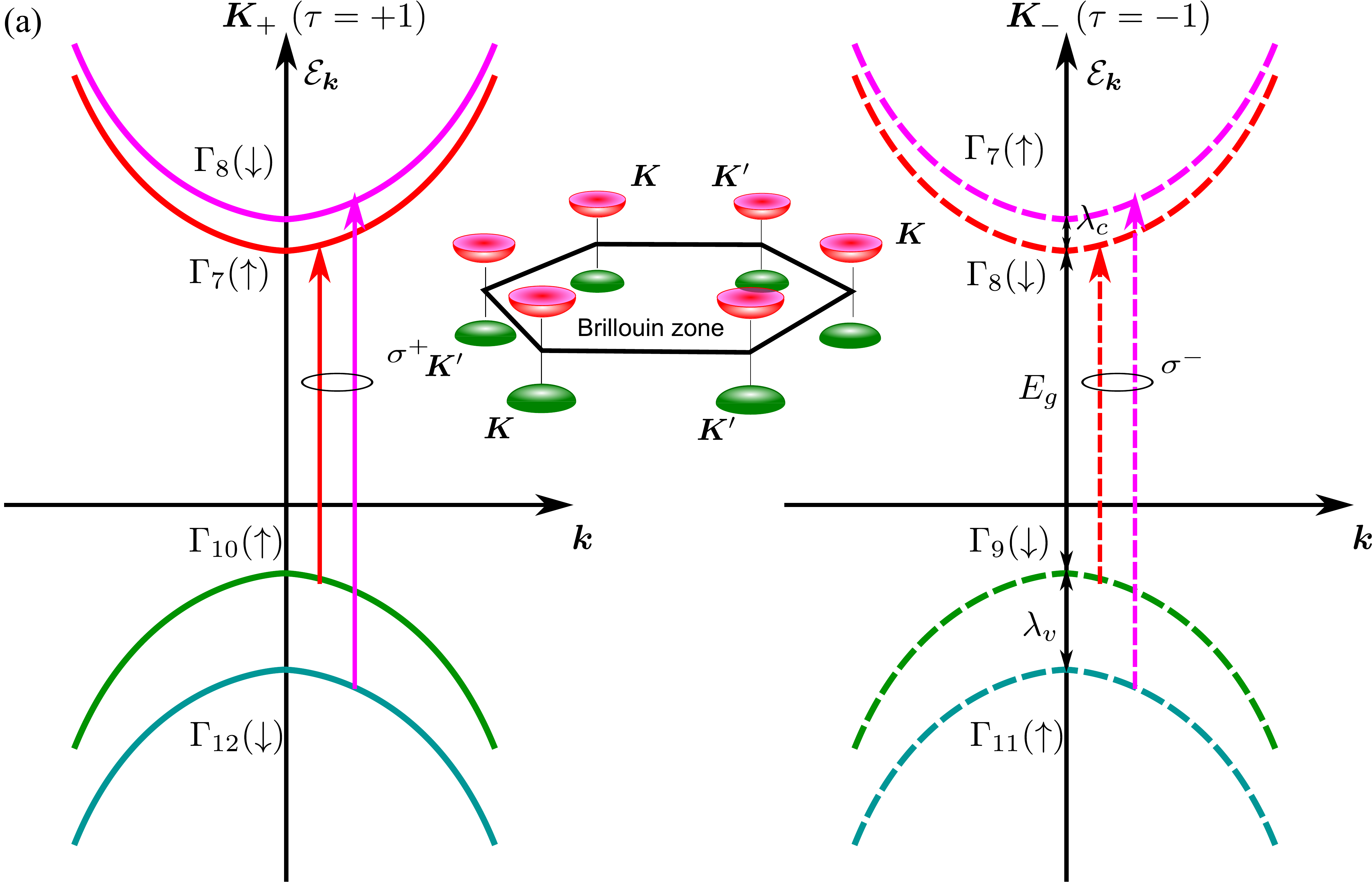}
\includegraphics[width=\linewidth]{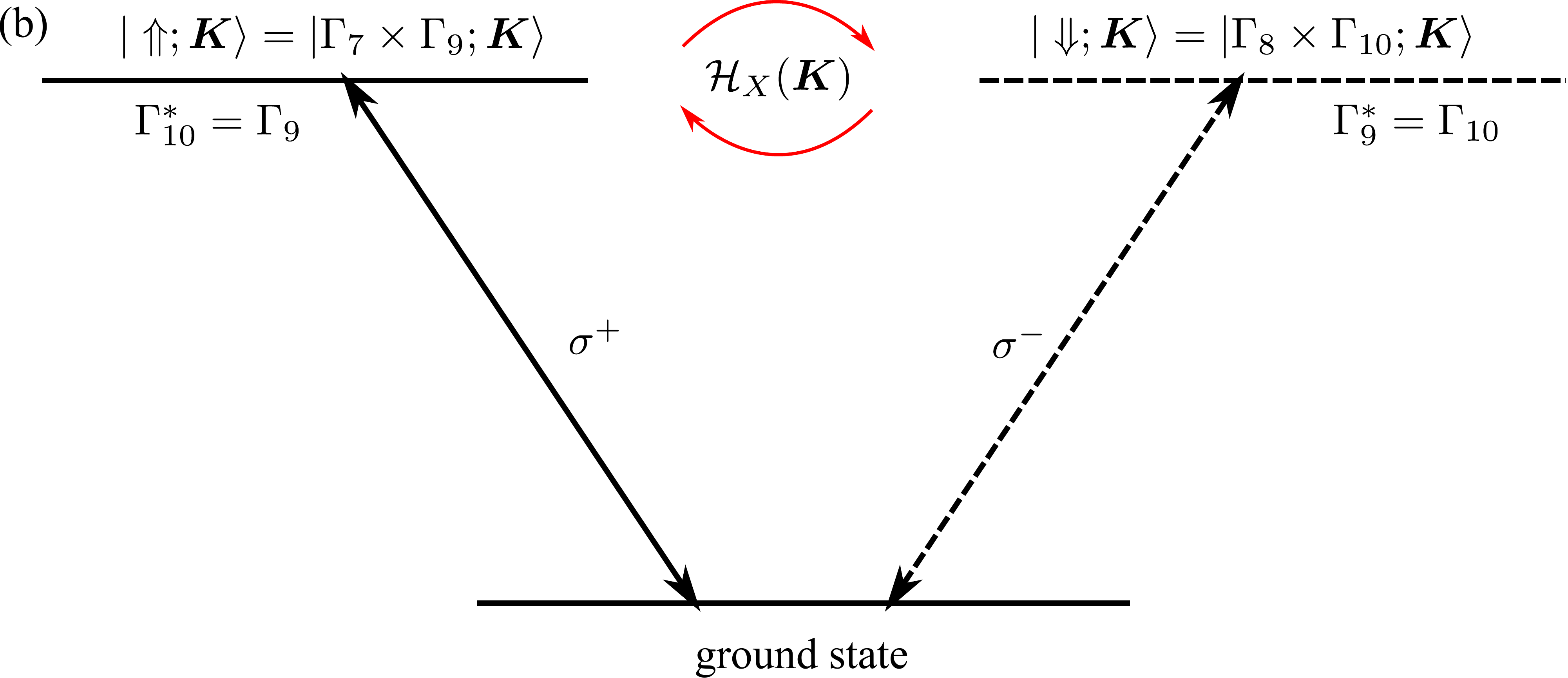}
\caption{(a) \textbf{Schematic illustration of MoS$_2$ band structure.} The bands are labelled by the corresponding irreducible representations with arrows in parentheses demonstrating electron spin orientation. Solid and dashed arrows show the transitions active at the normal incidence in $\sigma^+$ and $\sigma^-$ polarization, respectively. An inset sketches the Brillouin zone. The order of conduction band states is shown in accordance with Ref.~\onlinecite{Liu:2013if}. (b) \textbf{Optical selection rules of the two bright A-exciton states} with the small center of mass wavevector $\bm K$ and their long-range Coulomb exchange coupling, Eq.~\eqref{H:bright}. The $\Uparrow,\Downarrow$ symbols represent the exciton pseudospin in the reducible representation $\Gamma_2+\Gamma_3$. }\label{fig:bands}
\end{figure}

\textit{Model}. The point symmetry group of MoS$_2$-like dichalcogenide MLs is $D_{3h}$. Since the direct band gap is realized at the edges of hexagonal Brillouin zone, as shown in the inset to Fig.~\ref{fig:bands}, the symmetry of individual valley $\bm K_\pm$ is lower and is described by the $C_{3h}$ point group. The schematic band structure of ML MoS$_2$ is presented in Fig.~\ref{fig:bands}(a), where the bands are labelled according to the irreducible representations of $C_{3h}$ group in notations of Ref.~\onlinecite{koster63}. The analysis shows that all spinor representations of $C_{3h}$ are one-dimensional, therefore, in each valley all the states are non-degenerate. The Kramers degeneracy is recovered taking into account that the time reversal couples $\bm K_+$ and $\bm K_-$ valleys. 
The spin-orbit splitting between the valence subbands in a given valley ($\Gamma_{10}$ and $\Gamma_{12}$ at $\bm K_+$ point and $\Gamma_9$ and $\Gamma_{11}$ at $\bm K_-$ point)  is on the order of 100~meV.\cite{Xiao:2012cr} The conduction band in each valley is also split into two subbands $\Gamma_7$ and $\Gamma_8$  with the splitting being about several meV.\cite{Kormanyos:2013dq,Liu:2013if,Kosmider:2013a}

Within the $\bm k \cdot \bm p$ model we introduce basic Bloch functions $U_i^\tau(\bm r)$, which describe electron states at $\bm K_+$ ($\tau = 1$) or $\bm K_-$ ($\tau=-1$) point of the Brillouin zone transforming according to the irreducible representation $\Gamma_i$ of $C_{3h}$ point group. In the minimal approximation where contributions from distant bands are ignored, the Bloch amplitudes $U^\tau_{7}$ and $U^\tau_{8}$ of the conduction band states can be recast as products of spinors $\chi_{s_z}$ corresponding to the spin $z$ component, $s_z = \pm 1/2$, and orbital Bloch amplitudes $U^\tau_{1}$. Similarly, the valence band states in $\bm K_+$ valley, $U^{+1}_{{10}}$ and $U^{+1}_{{12}}$ are the products of $\chi_{s_z}$ and orbital function $U^{+1}_{3}$, while the valence band states in $\bm K_-$ valley, $U^{-1}_{11}$ and $U^{+1}_{{9}}$ originate from the orbital Bloch amplitude $U^{-1}_{2}$. Such a four-band $\bm k \cdot \bm p$ model is described by $4$ parameters: three energy gaps, namely, the band gap $E_g$, the spin splittings in the conduction and valence bands $\lambda_c$, $\lambda_v$, respectively, and interband momentum matrix element $p_{\rm cv} = \langle U^{+1}_1 | (p_x + \mathrm i p_y)/\sqrt{2}| U^{+1}_3\rangle = \langle U^{-1}_1 | (p_x - \mathrm i p_y)/\sqrt{2}| U^{-1}_2\rangle$. The latter definition is in agreement with remarkable optical selection rules:\cite{Cao:2012a,Mak:2012a,Zeng:2012ys} At a normal incidence of radiation the optical transitions in $\sigma^+$ polarization take place in $\bm K_+$ valley only between $\Gamma_{10}$ and $\Gamma_7$ or $\Gamma_{12}$ and $\Gamma_8$ states, while the transitions in $\sigma^-$ polarization take place in $\bm K_-$ valley and involve either $\Gamma_{9}$ and $\Gamma_8$ or $\Gamma_{11}$ and $\Gamma_7$ states. Each conduction subband is mixed by $\bm k \cdot \bm p$ interaction with the only valence subband having the same spin component. As an example we present below the wavefunctions obtained in the first order of $\bm k \cdot \bm p$ interaction for the bottom conduction subband~[cf.~Ref.~\onlinecite{birpikus_eng}]:
\begin{subequations}
\label{conduction}
\begin{align}
\psi_{c, \bm k}^{+1}(\bm r) & = \mathrm e^{\mathrm i \bm k \bm r} \left[U_{7,c}^{+1}(\bm r) + \frac{\hbar^2 p_{\rm cv}^* k_{+} }{m_0E_g} U_{10,v}^{+1}(\bm r) \right], \label{init:1}\\
\psi_{c, \bm k}^{-1}(\bm r) & = \mathrm e^{\mathrm i \bm k \bm r} \left[U_{8,c}^{-1}(\bm r) + \frac{\hbar^2 p_{\rm cv}^* k_{-} }{m_0E_g} U_{9,v}^{-1}(\bm r) \right],\label{fin:1}
\end{align}
\end{subequations}
and for the topmost valence subband
\begin{subequations}
\label{valence}
\begin{align}
\psi_{v,\bm k}^{+1}(\bm r)& = \mathrm e^{\mathrm i \bm k \bm r} \left[U_{10,v}^{+1}(\bm r) - \frac{\hbar^2 p_{\rm cv} k_{-}}{m_0E_g} U_{7,c}^{+1}(\bm r) \right],\label{fin:2}\\
\psi_{v,\bm k}^{-1}(\bm r) & = \mathrm e^{\mathrm i \bm k \bm r} \left[U_{9,v}^{-1}(\bm r) - \frac{\hbar^2 p_{\rm cv} k_{+}}{m_0E_g} U_{8,c}^{-1}(\bm r) \right].\label{init:2}
\end{align}
\end{subequations}
Here $\bm k$ is the wavevector reckoned from the $\bm K_+$ or $\bm K_-$ point, $k_{\pm} = (k_x \pm \mathrm i k_y)/\sqrt{2}$, asterisk denotes complex conjugate. The wavefunctions are taken in the electron representation and the factors $\exp{(\mathrm i \bm K_\pm \bm r)}$ are included in the definitions of Bloch amplitudes.

Optical excitation gives rise to the electron-hole pairs bound into excitons by the Coulomb interaction. Excitonic states transform according to the representations $\Gamma_c \times \Gamma_v^*$, where $\Gamma_c$ is the representation of the conduction band state and $\Gamma_v$ is the representation of the valence band state, and in $C_{3h}$ group the time reversed representation $\mathcal K\Gamma_v=\Gamma_v^*$. As a result, at normal incidence the optically active states are given by~[see Fig.~\ref{fig:bands}(b)]
\begin{subequations}
\label{exciton:repr}
\begin{align}
\Gamma_7 \times \Gamma_{10}^* = \Gamma_8 \times \Gamma_{12}^* = \Gamma_2\quad \mbox{active in }\sigma^+,\\
\Gamma_8 \times \Gamma_{9}^* = \Gamma_7 \times \Gamma_{11}^* = \Gamma_3\quad \mbox{active in }\sigma^-.
\end{align}
\end{subequations}
The aim of the paper is to study the fine structure of bright excitonic states and its consequences on the valley dynamics. Therefore, we do not address here the calculation of the ``mechanical'' exciton (the direct Coulomb problem),\cite{Cheiwchanchamnangij:2012pi,Qiu:2013fe} we assume that the relative electron-hole motion can be described by an envelope function $\varphi(\bm \rho)$.\footnote{The electrodynamical treatment holds also for the case of strongly bound excitons with the replacement of $e^2 |p_{\rm cv}|^2|\varphi(0)|^2/(m_0 \hbar\omega_0)$ in Eq.~\eqref{Gamma0} by exciton oscillator strength per unit area $f$.} The bright exciton fine structure can be calculated either within the $\bm k \cdot \bm p$ perturbation theory taking into account the long-range exchange interaction or by electrodynamical approach where the interaction of the mechanical exciton with generated electromagnetic field is taken into account. We start with the latter approach and then demonstrate its equivalence with the $\bm k \cdot \bm p$ calculation.

\begin{figure}[t]
\includegraphics[width=0.5\linewidth]{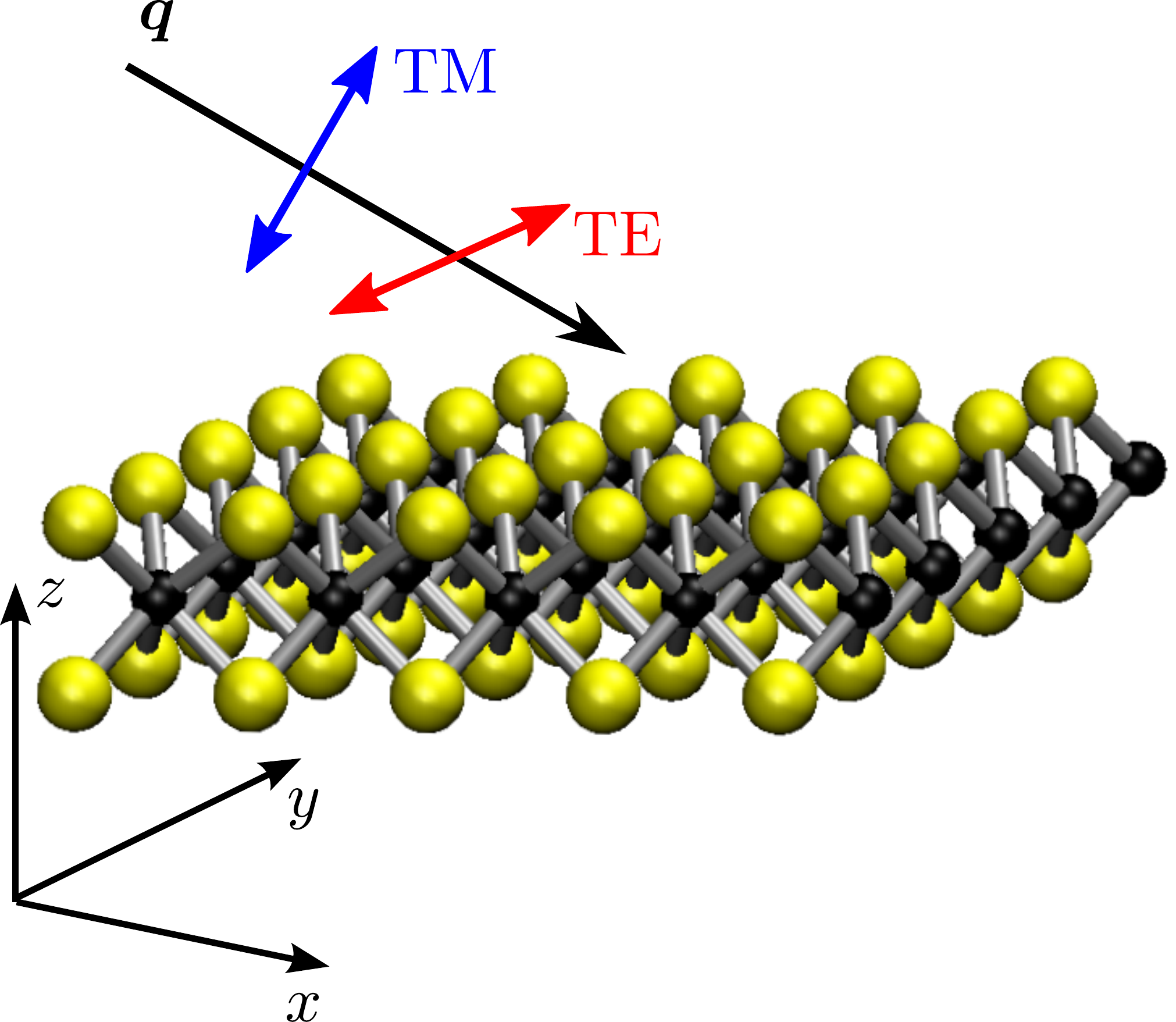}
\caption{\textbf{Schematic illustration the system geometry} with $s$ (TE mode) and $P$ (TM mode) polarized incident light.}\label{fig:geom}
\end{figure}

In the electrodynamic treatment the exciton frequencies can be found from the poles of the reflection coefficient of the two-dimensional structure.\cite{agr_ginz}
For simplicity we consider a ML of MoS$_2$ situated in $(xy)$ plane surrounded by dielectric media with high-frequency dielectric constant $\varkappa_b$, the contrast of background dielectric constants is disregarded.\footnote{In fact, $\varkappa_b$ should also include the contributions of transitions spectrally higher than the A-exciton in the MoS$_2$ layer. The allowance for the dielectric constant as well as for the substrate is straightforward following Ref.~\onlinecite{ivchenko05a}. It does not lead to substantial modifications of the results.} The geometry is illustrated in Fig.~\ref{fig:geom}.
We solve Maxwell equations for electromagnetic field taking into account the the excitonic contribution to the dielectric polarization, which in the vicinity of the A-exciton resonance reads
\begin{equation}
\label{nonloc}
\bm P(z) = \frac{ \delta(z) |\varphi(0)|^2 \bm E(z)}{\omega_0 - \omega + \mathrm i \Gamma} \frac{e^{2}|p_{\rm cv}|^2}{\hbar\omega_0^2m_0^2}.
\end{equation}
Here $\omega$ is the incident radiation frequency, $\omega_0$ is the exciton resonance frequency determined by the band gap and binding energy (its evaluation is beyond the scope of present paper), $\Gamma$ is its nonradiative damping. 
Factor $\delta(z)$ ensures that the dipole moment is induced only in the ML of MoS$_2$ whose width is negligible compared with the radiation wavelength. At a normal incidence of radiation the amplitude reflection coefficient of a ML has a standard form $r(\omega) = {\mathrm i \Gamma_0}/[{\omega_0 - \omega - \mathrm i (\Gamma_0+ \Gamma)}]$,\cite{ivchenko05a} 
where 
\begin{equation}
\label{Gamma0}
\Gamma_0 = \frac{2\pi q e^2 |p_{\rm cv}|^2}{\hbar\varkappa_b \omega_0^2m_0^2} {|\varphi(0)|^2}, 
\end{equation}
is the radiative decay rate of an exciton in the MoS$_2$ ML, $q= \sqrt{\varkappa_b} \omega/c$ is the wavevector of radiation. The parameters of the pole in the reflectivity describe the eigenenergy and decay rate of the exciton with allowance for the light-matter interaction.
In agreement with symmetry, the reflection coefficient at a normal incidence is polarization independent. 

Under oblique incidence in the $(xz)$-plane the solution of Maxwell equation yields two eigenmodes of electromagnetic field: $s$-polarized (TE-polarized) wave with $\bm E\parallel y$ (perpendicular to the light incidence plane) and $p$-polarized (TM-polarized) wave with $\bm E$ in the incidence plane, see Fig.~\ref{fig:geom}. The reflection coefficients in a given polarization $\alpha=s$ or $p$ are given by the pole contributions with the modified parameters
\begin{equation}
\label{r:alpha}
r_\alpha(\omega) = \frac{\mathrm i \Gamma_{0\alpha}}{\omega_{0\alpha} - \omega - \mathrm i (\Gamma_{0\alpha}+ \Gamma)}, 
\end{equation}
where
\begin{equation}
\label{Gamma0s}
\Gamma_{0s} = \frac{q}{q_z} \Gamma_0, \quad \Gamma_{0p} = \frac{q_z}{q} \Gamma_0,
\end{equation}
$q_z = (q^2 - q_\parallel^2)^{1/2}$ is the $z$ component of the light wavevector, $q_\parallel$ is its in-plane component, and $\omega_{0\alpha} \equiv \omega_0(q_\parallel)= \omega_0 + \hbar q_\parallel^2/(2M)$ is the mechanical exciton frequency, $M$ is the exciton effective mass.\footnote{Note, that at $q_\parallel \ne 0$ the transitions $\Gamma_{10}\to \Gamma_{8}$ and $\Gamma_{9} \to \Gamma_{7}$ active in $z$ polarization become symmetry-allowed (but requiring a spin flip). It gives rise to an additional pole in $r_p(\omega)$ at the frequency $\omega_0' \approx \omega_0 + \lambda_c/\hbar$ and can be evaluated following Ref.~\onlinecite{ivchenko05a}.}

The light-matter interaction results in the radiative decay of the excitons with the wavevectors inside the light cone, $q_\parallel \leqslant \sqrt{\varkappa_b} \omega/c$. For the excitons outside the light cone, $q_z$ becomes imaginary and corresponding exciton-induced electromagnetic field decays exponentially with the distance from the ML. Therefore, exciton interaction with the field results in renormalization of its frequency rather than its decay rate.\cite{goupalov98,ivchenko05a,goupalov:electrodyn} Formally, it corresponds to imaginary $\Gamma_{0\alpha}$ in Eqs.~\eqref{Gamma0s}. 
Introducing the notation $\bm K = \bm q_{\parallel}$ for the center of mass wavevector of an exciton, we obtain from the poles of reflection coefficients the splitting between the longitudinal ($\bm P \parallel \bm K$) and transverse ($\bm P \perp \bm K$) exciton states:
\begin{equation}
\label{LTsplitting}
\Delta E = \hbar \Gamma_0 \frac{K^2}{q \sqrt{K^2 - q^2}} \approx \hbar \Gamma_0 \frac{K}{q},
\end{equation}
where the approximate equation holds for $K\gg q$ and one can replace $\omega$ by $\omega_0(K)$ or even by $\omega_0(0)$ in the definition of $q$.
 For excitons outside the light cone and arbitrary in-plane direction of $\bm K$ one can present an effective Hamiltonian describing their fine structure in the basis of states $\Gamma_2$ and $\Gamma_3$ [see Eqs.~\eqref{exciton:repr}] as
\begin{equation}
\label{H:bright}
\mathcal H_{X} (\bm K)=  
\begin{pmatrix}
0 & \alpha (K_x - \mathrm i K_y)^2 \\
 \alpha  (K_x+ \mathrm i K_y)^2 & 0
\end{pmatrix} = \frac{\hbar}{2} \left( \bm \Omega_{\bm K} \cdot \bm \sigma \right).
\end{equation}
Here $\alpha = \hbar \Gamma_0/(2 K q)$, $\bm \sigma = (\sigma_x,\sigma_y)$ are two pseudospin Pauli matrices and the effective spin precession frequency vector has the following components $\hbar \Omega_{x} = \Delta E \cos{2\vartheta}$ and $\hbar \Omega_{y} = \Delta E \sin{2\vartheta}$, where $\vartheta$ is the angle between $\bm K$ and the in-plane axis $x$.

Now we demonstrate that the treatment based on electrodynamics presented above is equivalent to the $\bm k\cdot \bm p$ calculation of the long-range exchange interaction. Specifically, we consider the exchange interaction between two electrons $\psi_m$ and $\psi_n$ occupying $\Gamma_7$ band in $\bm K_+$ valley and $\Gamma_9$ electron in $\bm K_-$ valley. The wavefunctions of these states are given by Eqs.~\eqref{init:1} and \eqref{init:2} with the wavevectors $\bm k_1$ and $\bm k_2$, respectively. The final states for the pair are the conduction band $\Gamma_8$ state in $\bm K_-$ valley, $\psi_{m'}$, and valence band  $\Gamma_{10}$ state in $\bm K_+$ valley, $\psi_{n'}$ characterized by the wavevectors $\bm k_1'$ and $\bm k_2'$ and described by the wavefunctions Eqs.~\eqref{fin:1}, \eqref{fin:2}, respectively.  
 According to the general theory\cite{birpikus_eng} the exchange matrix element of the Coulomb interaction $V(\bm r_1 - \bm r_2)$ can be written as
\begin{multline}
\label{exchange}
\langle m' n'| V(\bm r_1 - \bm r_2)| m n\rangle = -V_{\bm k_1' - \bm k_2} \delta_{\bm k_1 + \bm k_2, \bm k_1'+\bm k_2'} \times \\ {\frac{\hbar^2 |p_{\rm cv}|^2}{m_0^2 E_g^2}}(k_{2,+} - k_{2,+}')(k_{1,+}- k_{1,+}').
\end{multline}
Here $V_{\bm q}$ is the two-dimensional Fourier transform of the Coulomb potential.  Standard transformations from the electron-electron representation to the electron-hole representation in Eq.~\eqref{exchange},\cite{birpikus_eng} averaging over the relative motion wavefunction as well as inclusion of high-frequency screening (see Refs.~\onlinecite{goupalov98,zhilich72:eng,zhilich74:eng}) yields off-diagonal element $\langle \Gamma_3 |\mathcal H_X (\bm K)|\Gamma_2 \rangle$ in Eq.~\eqref{H:bright}. We stress that the Coulomb interaction is long-range, it does not provide intervalley transfer of electrons, however, the exchange process involves one electron from $\bm K_+$ and another one from $\bm K_-$ valley. As a result, states active in $\sigma^+$ and $\sigma^-$ polarizations can be mixed (this fact was ignored in Ref.~\onlinecite{Yu:2014uq}).
%

In order to evaluate theoretically the exciton decoherence rate it is convenient to describe the dynamics of bright exciton doublet within the pseudospin formalism, where the $2\times 2$ density matrix in the basis of $\Gamma_2$ and $\Gamma_3$ excitonic states is decomposed $N_{\bm K}/2 + \bm S_{\bm K} \cdot \bm \sigma$ with $N_{\bm K}$ being the occupation of a given $\bm K$ state and $\bm S_{\bm K}$ being the pseudospin. It satisfies the kinetic equation
\begin{equation}
\label{kin}
\frac{\partial \bm S_{\bm K}}{\partial t} + \bm S_{\bm K} \times \bm \Omega_{\bm K} = \bm Q\{ \bm S_{\bm K} \},
\end{equation}
where $\bm \Omega_{\bm K}$ is defined by Eq.~\eqref{H:bright} and $\bm Q\{ \bm S_{\bm K} \}$ is the collision integral. Similarly to the case of free excitons in quantum wells\cite{maialle93,ivchenko05a} different regimes of spin decoherence can be identified depending on the relation between the characteristic spin precession frequency and scattering rates. 
Here we assume the strong scattering regime, where $\Omega \tau \ll 1$, where $\Omega$ is the characteristic precession frequency and $\tau$ is the characteristic scattering time, the exciton spin is lost by Dyakonov-Perel' type mechanism.\cite{dyakonov72} Hence, the spin decay law is exponential and spin relaxation rates  are given by\cite{maialle93,ivchenko05a}
\begin{equation}
\label{tau:s:scatt}
\frac{1}{\tau_{zz}} = \frac{2}{\tau_{xx}} = \frac{2}{\tau_{yy}} = \langle \Omega_{\bm K}^2 \tau_2\rangle,
\end{equation}
where the angular brackets denote averaging over the energy distribution and $\tau_2 = \tau_2(\varepsilon_{\bm K})$ is the relaxation time of second angular harmonics  of the distribution function. 

\begin{figure}
\includegraphics[width=0.8\linewidth]{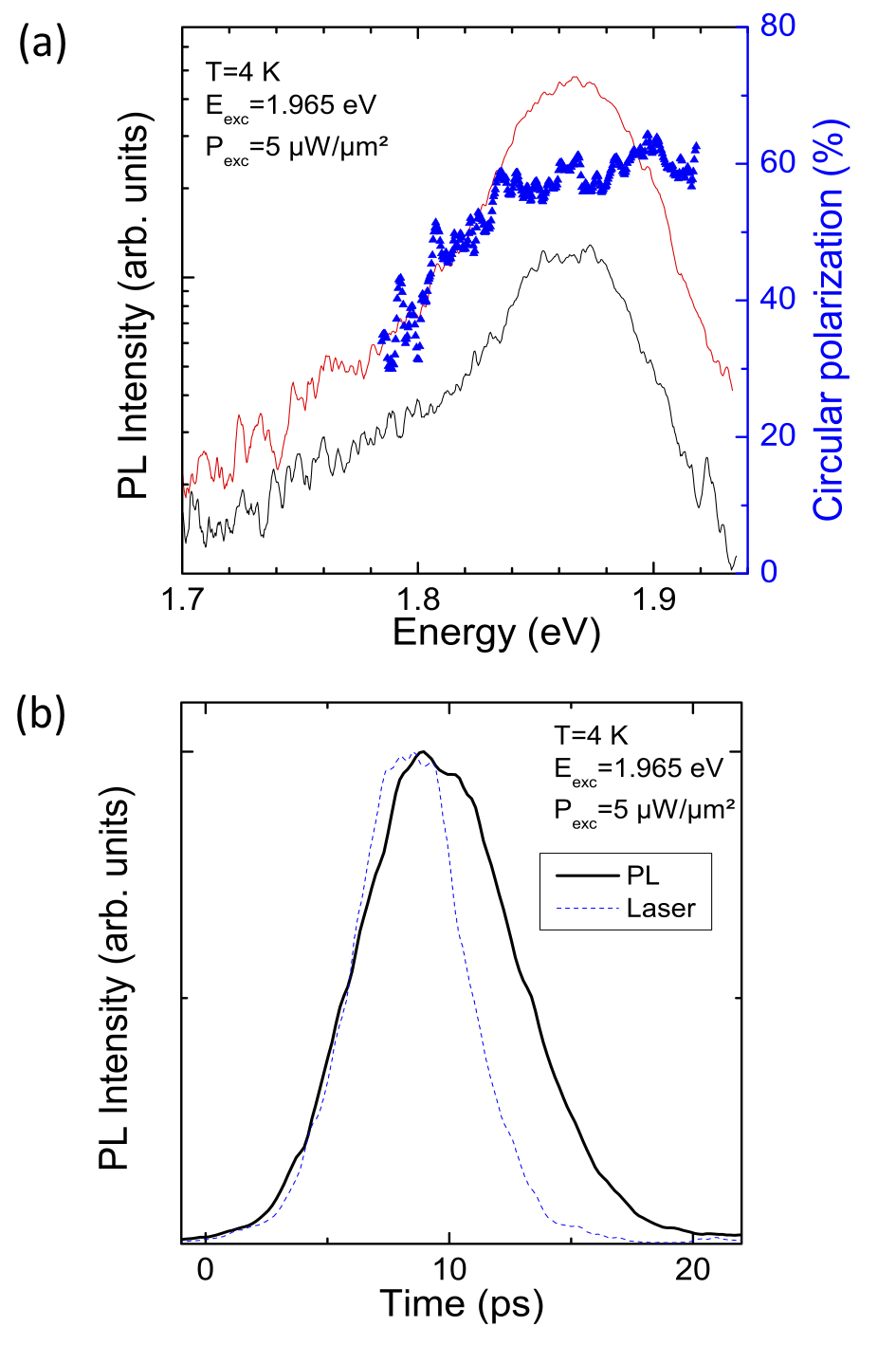}
\caption{\label{fig:TRPL} \textbf{Photoluminescence experiments on A-exciton in ML MoS$_2$.} (a) Left axis: Time integrated PL intensity as a function of emission energy. Right axis: Polarization of PL emission (b) PL emission intensity (black line) at T = 4~K detected at maximum of A-exciton PL E$_{\text{Det}}=1.867$~eV as a function of time. Laser reference pulse (dotted blue line).}
\end{figure}

\textit{Experiment and discussion}. In this section we discuss exciton spin or valley decoherence due to the long-range exchange interaction and compare theoretical estimates with our experimental data. Figure~\ref{fig:TRPL} presents the results of photoluminescence (PL) experiments carried out on MoS$_2$ ML deposited on the SiO$_2$/Si substrate, see Ref.~\onlinecite{PhysRevLett.112.047401} for details on sample preparation and experimental methods. The sample was excited by short circularly polarized laser pulse with the energy $E_{exc} = 1.965$~eV. Both PL intensity and circular polarization degree were recorded as a function of emission energy. As a simplest possible model, we assume that the stationary (time integrated) polarization is determined by the initially created polarization $P_0$, the lifetime of the electron-hole pair $\tau$ and the polarization decay time $\tau_{zz}$ through $P_c=P_0/(1+\tau/\tau_{zz})$.\cite{opt_or_book} In Fig.~\ref{fig:TRPL}(a) we measure an average, time-integrated PL polarization of $P_c\approx 60\%$ in the emission energy range $1.82\ldots 1.91$~eV. The emission time measured in time-resolved PL is $\tau \simeq 4.5$~ps, extracted from Fig.~\ref{fig:TRPL}(b), see also Ref.~\onlinecite{KornMoS2}. Note, that it is not clear at this stage if the measured emission time is an intrinsic, radiative lifetime or limited by non-radiative processes. For $P_0=100\%$ we find an  estimate of $\tau_{zz}\simeq 7$~ps. 

A theoretical estimate of $\tau_{zz}$ can be obtained taking into account that, in our experimental conditions, owing to the fast energy relaxation in the system, the spread of excitons in the energy space is limited by the collisional broadening, $\sim \hbar/\tau_2$, rather than by the kinetic energy distribution. Under this assumption, the trend for the spin decoherence rate can be obtained from Eqs.~\eqref{tau:s:scatt} taking into account that~[cf. Ref.~\onlinecite{maialle93}]
\begin{equation}
\label{eq:Est1}
\langle \Omega_{\bm K}^2 \tau_2 \rangle\simeq \Omega_{\bm K_\Gamma}^2 \tau_2,
\end{equation}
where $K_\Gamma = \sqrt{{2M\Gamma_h}/{\hbar^2}}$ and $\Gamma_h=1/(2\tau_2)$ describes the $\bm k$-space extension of an excitonic ``packet''. It follows then that the scattering time cancels in the right hand side of Eq.~\eqref{eq:Est1} and 
\begin{equation}
\label{eq:tau_zz}
 \tau_{zz} = \frac{4\hbar  \left(q \tau_{rad}\right)^2}{M}
\end{equation}
For an order of magnitude estimate, we set exciton mass $M$ equal to the free electron mass.\cite{Cheiwchanchamnangij:2012pi} We also set $\tau \approx \tau_{rad} = 4.5$~ps. Assuming the PL decay time is governed by radiative processes, realistically $\tau$ should be regarded as a lower bound of $\tau_{rad}$. We estimate the radiation wavevector $q= \sqrt{\varkappa_b}\omega_0/c$ assuming $\hbar\omega_0=1.867$~eV and $\varkappa_b=5$ (being half of the substrate high-frequency dielectric constant), and obtain $\tau_{zz}\approx 4$~ps. This value is in reasonable agreement with the value of $\tau_s\simeq 7$~ps estimated from PL experiments and with recent pump-probe measurements.\cite{doi:10.1021/nl403742j,Wang:2013hb}

Similarly to the circular polarization degree whose decay is governed by $\tau_{zz}$, the linear polarization decay for the neutral A-exciton is governed by the in-plane pseudospin relaxation times $\tau_{xx}$ and $\tau_{yy}$. As follows from Eq.~\eqref{tau:s:scatt} they are of the same order of magnitude as $\tau_{zz}$. Interestingly, the observation of linearly polarized emission under the linearly polarized excitation was reported for the neutral A-exciton transition in ML WSe$_2$.\cite{Jones:2013tg}  Since the band structure and parameters of WSe$_2$ and MoS$_2$ are similar, it is reasonable to assume that the decay of linear polarization, i.e. intervalley coherence of excitons, is also strongly influenced by the long-range exchange interaction between an electron and a hole.


\textit{Conclusions.} To conclude we have presented the theory of the bright exciton fine structure splitting and exciton spin decoherence in MLs of transition metal dichalcogenides. Using the electrodynamical approach we have calculated eigenfrequencies of excitons taking into account their interaction with longitudinal electromagnetic field, which gives rise to the LT splitting of the bright excitonic doublet. The magnitude of the splitting is expressed via the exciton center of mass wavevector and its radiative decay rate $\Gamma_0$. This splitting acts as an effective magnetic field and provides spin relaxation/decoherence of both free and localized excitons. We provided estimation of spin decoherence rate of A-excitons in MoS$_2$ MLs both from the developed theory and from experimental data on optical orientation. The developed theory is in agreement with experiments probing the exciton valley dynamics.

\textit{Acknowledgements.} We thank E.L.~Ivchenko and B.L.~Liu for stimulating discussions. Partial financial support from RFBR and RF President grant NSh-1085.2014.2, INSA invited Professorship grant (MMG), ERC Starting Grant No. 306719 and Programme Investissements d'Avenir ANR-11-IDEX-0002-02, reference ANR-10-LABX-0037-NEXT is gratefully acknowledged. 

\bibliography{mos2bibBU}

\begin{thebibliography}{37}
\expandafter\ifx\csname natexlab\endcsname\relax\def\natexlab#1{#1}\fi
\expandafter\ifx\csname bibnamefont\endcsname\relax
  \def\bibnamefont#1{#1}\fi
\expandafter\ifx\csname bibfnamefont\endcsname\relax
  \def\bibfnamefont#1{#1}\fi
\expandafter\ifx\csname citenamefont\endcsname\relax
  \def\citenamefont#1{#1}\fi
\expandafter\ifx\csname url\endcsname\relax
  \def\url#1{\texttt{#1}}\fi
\expandafter\ifx\csname urlprefix\endcsname\relax\def\urlprefix{URL }\fi
\providecommand{\bibinfo}[2]{#2}
\providecommand{\eprint}[2][]{\url{#2}}

\bibitem[{\citenamefont{Xiao et~al.}(2012)\citenamefont{Xiao, Liu, Feng, Xu,
  and Yao}}]{Xiao:2012cr}
\bibinfo{author}{\bibfnamefont{D.}~\bibnamefont{Xiao}},
  \bibinfo{author}{\bibfnamefont{G.-B.} \bibnamefont{Liu}},
  \bibinfo{author}{\bibfnamefont{W.}~\bibnamefont{Feng}},
  \bibinfo{author}{\bibfnamefont{X.}~\bibnamefont{Xu}}, \bibnamefont{and}
  \bibinfo{author}{\bibfnamefont{W.}~\bibnamefont{Yao}},
  \bibinfo{journal}{Phys. Rev. Lett.} \textbf{\bibinfo{volume}{108}},
  \bibinfo{pages}{196802} (\bibinfo{year}{2012}).

\bibitem[{\citenamefont{Cao et~al.}(2012)\citenamefont{Cao, Wang, Han, Ye, Zhu,
  Shi, Niu, Tan, Wang, Liu et~al.}}]{Cao:2012a}
\bibinfo{author}{\bibfnamefont{T.}~\bibnamefont{Cao}},
  \bibinfo{author}{\bibfnamefont{G.}~\bibnamefont{Wang}},
  \bibinfo{author}{\bibfnamefont{W.}~\bibnamefont{Han}},
  \bibinfo{author}{\bibfnamefont{H.}~\bibnamefont{Ye}},
  \bibinfo{author}{\bibfnamefont{C.}~\bibnamefont{Zhu}},
  \bibinfo{author}{\bibfnamefont{J.}~\bibnamefont{Shi}},
  \bibinfo{author}{\bibfnamefont{Q.}~\bibnamefont{Niu}},
  \bibinfo{author}{\bibfnamefont{P.}~\bibnamefont{Tan}},
  \bibinfo{author}{\bibfnamefont{E.}~\bibnamefont{Wang}},
  \bibinfo{author}{\bibfnamefont{B.}~\bibnamefont{Liu}}, \bibnamefont{et~al.},
  \bibinfo{journal}{Nature Communications} \textbf{\bibinfo{volume}{3}},
  \bibinfo{pages}{887} (\bibinfo{year}{2012}).

\bibitem[{\citenamefont{Mak et~al.}(2012)\citenamefont{Mak, He, Shan, and
  Heinz}}]{Mak:2012a}
\bibinfo{author}{\bibfnamefont{K.~F.} \bibnamefont{Mak}},
  \bibinfo{author}{\bibfnamefont{K.}~\bibnamefont{He}},
  \bibinfo{author}{\bibfnamefont{J.}~\bibnamefont{Shan}}, \bibnamefont{and}
  \bibinfo{author}{\bibfnamefont{T.~F.} \bibnamefont{Heinz}},
  \bibinfo{journal}{Nat. Nanotechnol.} \textbf{\bibinfo{volume}{7}},
  \bibinfo{pages}{494} (\bibinfo{year}{2012}).

\bibitem[{\citenamefont{Zeng et~al.}(2012)\citenamefont{Zeng, Dai, Yao, Xiao,
  and Cui}}]{Zeng:2012ys}
\bibinfo{author}{\bibfnamefont{H.}~\bibnamefont{Zeng}},
  \bibinfo{author}{\bibfnamefont{J.}~\bibnamefont{Dai}},
  \bibinfo{author}{\bibfnamefont{W.}~\bibnamefont{Yao}},
  \bibinfo{author}{\bibfnamefont{D.}~\bibnamefont{Xiao}}, \bibnamefont{and}
  \bibinfo{author}{\bibfnamefont{X.}~\bibnamefont{Cui}}, \bibinfo{journal}{Nat
  Nano} \textbf{\bibinfo{volume}{7}}, \bibinfo{pages}{490}
  (\bibinfo{year}{2012}).

\bibitem[{\citenamefont{Sallen et~al.}(2012)\citenamefont{Sallen, Bouet, Marie,
  Wang, Zhu, Han, Lu, Tan, Amand, Liu et~al.}}]{Sallen:2012qf}
\bibinfo{author}{\bibfnamefont{G.}~\bibnamefont{Sallen}},
  \bibinfo{author}{\bibfnamefont{L.}~\bibnamefont{Bouet}},
  \bibinfo{author}{\bibfnamefont{X.}~\bibnamefont{Marie}},
  \bibinfo{author}{\bibfnamefont{G.}~\bibnamefont{Wang}},
  \bibinfo{author}{\bibfnamefont{C.~R.} \bibnamefont{Zhu}},
  \bibinfo{author}{\bibfnamefont{W.~P.} \bibnamefont{Han}},
  \bibinfo{author}{\bibfnamefont{Y.}~\bibnamefont{Lu}},
  \bibinfo{author}{\bibfnamefont{P.~H.} \bibnamefont{Tan}},
  \bibinfo{author}{\bibfnamefont{T.}~\bibnamefont{Amand}},
  \bibinfo{author}{\bibfnamefont{B.~L.} \bibnamefont{Liu}},
  \bibnamefont{et~al.}, \bibinfo{journal}{Phys. Rev. B}
  \textbf{\bibinfo{volume}{86}}, \bibinfo{pages}{081301}
  (\bibinfo{year}{2012}).

\bibitem[{\citenamefont{Kioseoglou et~al.}(2012)\citenamefont{Kioseoglou,
  Hanbicki, Currie, Friedman, Gunlycke, and Jonker}}]{Kioseoglou}
\bibinfo{author}{\bibfnamefont{G.}~\bibnamefont{Kioseoglou}},
  \bibinfo{author}{\bibfnamefont{A.~T.} \bibnamefont{Hanbicki}},
  \bibinfo{author}{\bibfnamefont{M.}~\bibnamefont{Currie}},
  \bibinfo{author}{\bibfnamefont{A.~L.} \bibnamefont{Friedman}},
  \bibinfo{author}{\bibfnamefont{D.}~\bibnamefont{Gunlycke}}, \bibnamefont{and}
  \bibinfo{author}{\bibfnamefont{B.~T.} \bibnamefont{Jonker}},
  \bibinfo{journal}{Applied Physics Letters} \textbf{\bibinfo{volume}{101}},
  \bibinfo{eid}{221907} (\bibinfo{year}{2012}).

\bibitem[{\citenamefont{Lagarde et~al.}(2014)\citenamefont{Lagarde, Bouet,
  Marie, Zhu, Liu, Amand, Tan, and Urbaszek}}]{PhysRevLett.112.047401}
\bibinfo{author}{\bibfnamefont{D.}~\bibnamefont{Lagarde}},
  \bibinfo{author}{\bibfnamefont{L.}~\bibnamefont{Bouet}},
  \bibinfo{author}{\bibfnamefont{X.}~\bibnamefont{Marie}},
  \bibinfo{author}{\bibfnamefont{C.~R.} \bibnamefont{Zhu}},
  \bibinfo{author}{\bibfnamefont{B.~L.} \bibnamefont{Liu}},
  \bibinfo{author}{\bibfnamefont{T.}~\bibnamefont{Amand}},
  \bibinfo{author}{\bibfnamefont{P.~H.} \bibnamefont{Tan}}, \bibnamefont{and}
  \bibinfo{author}{\bibfnamefont{B.}~\bibnamefont{Urbaszek}},
  \bibinfo{journal}{Phys. Rev. Lett.} \textbf{\bibinfo{volume}{112}},
  \bibinfo{pages}{047401} (\bibinfo{year}{2014}).

\bibitem[{\citenamefont{Zhu et~al.}(2013)\citenamefont{Zhu, Wang, Liu, Marie,
  Qiao, Zhang, Wu, Fan, Tan, Amand et~al.}}]{Zhu:2013ve}
\bibinfo{author}{\bibfnamefont{C.~R.} \bibnamefont{Zhu}},
  \bibinfo{author}{\bibfnamefont{G.}~\bibnamefont{Wang}},
  \bibinfo{author}{\bibfnamefont{B.~L.} \bibnamefont{Liu}},
  \bibinfo{author}{\bibfnamefont{X.}~\bibnamefont{Marie}},
  \bibinfo{author}{\bibfnamefont{X.~F.} \bibnamefont{Qiao}},
  \bibinfo{author}{\bibfnamefont{X.}~\bibnamefont{Zhang}},
  \bibinfo{author}{\bibfnamefont{X.~X.} \bibnamefont{Wu}},
  \bibinfo{author}{\bibfnamefont{H.}~\bibnamefont{Fan}},
  \bibinfo{author}{\bibfnamefont{P.~H.} \bibnamefont{Tan}},
  \bibinfo{author}{\bibfnamefont{T.}~\bibnamefont{Amand}},
  \bibnamefont{et~al.}, \bibinfo{journal}{Phys. Rev. B}
  \textbf{\bibinfo{volume}{88}}, \bibinfo{pages}{121301}
  (\bibinfo{year}{2013}).

\bibitem[{\citenamefont{Li et~al.}(2013)\citenamefont{Li, Zhang, and
  Niu}}]{Li:2013qf}
\bibinfo{author}{\bibfnamefont{X.}~\bibnamefont{Li}},
  \bibinfo{author}{\bibfnamefont{F.}~\bibnamefont{Zhang}}, \bibnamefont{and}
  \bibinfo{author}{\bibfnamefont{Q.}~\bibnamefont{Niu}},
  \bibinfo{journal}{Phys. Rev. Lett.} \textbf{\bibinfo{volume}{110}},
  \bibinfo{pages}{066803} (\bibinfo{year}{2013}).

\bibitem[{\citenamefont{Molina-S\'anchez
  et~al.}(2013)\citenamefont{Molina-S\'anchez, Sangalli, Hummer, Marini, and
  Wirtz}}]{Molina-Sanchez:2013mi}
\bibinfo{author}{\bibfnamefont{A.}~\bibnamefont{Molina-S\'anchez}},
  \bibinfo{author}{\bibfnamefont{D.}~\bibnamefont{Sangalli}},
  \bibinfo{author}{\bibfnamefont{K.}~\bibnamefont{Hummer}},
  \bibinfo{author}{\bibfnamefont{A.}~\bibnamefont{Marini}}, \bibnamefont{and}
  \bibinfo{author}{\bibfnamefont{L.}~\bibnamefont{Wirtz}},
  \bibinfo{journal}{Phys. Rev. B} \textbf{\bibinfo{volume}{88}},
  \bibinfo{pages}{045412} (\bibinfo{year}{2013}).

\bibitem[{\citenamefont{Ochoa and Rold\'an}(2013)}]{Ochoa:2013wd}
\bibinfo{author}{\bibfnamefont{H.}~\bibnamefont{Ochoa}} \bibnamefont{and}
  \bibinfo{author}{\bibfnamefont{R.}~\bibnamefont{Rold\'an}},
  \bibinfo{journal}{Phys. Rev. B} \textbf{\bibinfo{volume}{87}},
  \bibinfo{pages}{245421} (\bibinfo{year}{2013}).

\bibitem[{\citenamefont{Wang and Wu}(2013)}]{MWWuMoS2}
\bibinfo{author}{\bibfnamefont{L.}~\bibnamefont{Wang}} \bibnamefont{and}
  \bibinfo{author}{\bibfnamefont{M.~W.} \bibnamefont{Wu}},
  \bibinfo{journal}{Preprint arXiv:1305.3361}  (\bibinfo{year}{2013}).

\bibitem[{\citenamefont{Song and Dery}(2013)}]{Song:2013uq}
\bibinfo{author}{\bibfnamefont{Y.}~\bibnamefont{Song}} \bibnamefont{and}
  \bibinfo{author}{\bibfnamefont{H.}~\bibnamefont{Dery}},
  \bibinfo{journal}{Phys. Rev. Lett.} \textbf{\bibinfo{volume}{111}},
  \bibinfo{pages}{026601} (\bibinfo{year}{2013}).

\bibitem[{\citenamefont{Mai et~al.}(2014)\citenamefont{Mai, Barrette, Yu,
  Semenov, Kim, Cao, and Gundogdu}}]{doi:10.1021/nl403742j}
\bibinfo{author}{\bibfnamefont{C.}~\bibnamefont{Mai}},
  \bibinfo{author}{\bibfnamefont{A.}~\bibnamefont{Barrette}},
  \bibinfo{author}{\bibfnamefont{Y.}~\bibnamefont{Yu}},
  \bibinfo{author}{\bibfnamefont{Y.~G.} \bibnamefont{Semenov}},
  \bibinfo{author}{\bibfnamefont{K.~W.} \bibnamefont{Kim}},
  \bibinfo{author}{\bibfnamefont{L.}~\bibnamefont{Cao}}, \bibnamefont{and}
  \bibinfo{author}{\bibfnamefont{K.}~\bibnamefont{Gundogdu}},
  \bibinfo{journal}{Nano Letters} \textbf{\bibinfo{volume}{14}},
  \bibinfo{pages}{202} (\bibinfo{year}{2014}).

\bibitem[{\citenamefont{Wang et~al.}(2013)\citenamefont{Wang, Ge, Li, Qiu, Ji,
  Feng, and Sun}}]{Wang:2013hb}
\bibinfo{author}{\bibfnamefont{Q.}~\bibnamefont{Wang}},
  \bibinfo{author}{\bibfnamefont{S.}~\bibnamefont{Ge}},
  \bibinfo{author}{\bibfnamefont{X.}~\bibnamefont{Li}},
  \bibinfo{author}{\bibfnamefont{J.}~\bibnamefont{Qiu}},
  \bibinfo{author}{\bibfnamefont{Y.}~\bibnamefont{Ji}},
  \bibinfo{author}{\bibfnamefont{J.}~\bibnamefont{Feng}}, \bibnamefont{and}
  \bibinfo{author}{\bibfnamefont{D.}~\bibnamefont{Sun}}, \bibinfo{journal}{ACS
  Nano} \textbf{\bibinfo{volume}{7}}, \bibinfo{pages}{11087}
  (\bibinfo{year}{2013}).

\bibitem[{\citenamefont{Cheiwchanchamnangij and
  Lambrecht}(2012)}]{Cheiwchanchamnangij:2012pi}
\bibinfo{author}{\bibfnamefont{T.}~\bibnamefont{Cheiwchanchamnangij}}
  \bibnamefont{and} \bibinfo{author}{\bibfnamefont{W.~R.~L.}
  \bibnamefont{Lambrecht}}, \bibinfo{journal}{Phys. Rev. B}
  \textbf{\bibinfo{volume}{85}}, \bibinfo{pages}{205302}
  (\bibinfo{year}{2012}).

\bibitem[{\citenamefont{Mak et~al.}(2013)\citenamefont{Mak, He, Lee, Lee, Hone,
  Heinz, and Shan}}]{Mak:2013lh}
\bibinfo{author}{\bibfnamefont{K.~F.} \bibnamefont{Mak}},
  \bibinfo{author}{\bibfnamefont{K.}~\bibnamefont{He}},
  \bibinfo{author}{\bibfnamefont{C.}~\bibnamefont{Lee}},
  \bibinfo{author}{\bibfnamefont{G.~H.} \bibnamefont{Lee}},
  \bibinfo{author}{\bibfnamefont{J.}~\bibnamefont{Hone}},
  \bibinfo{author}{\bibfnamefont{T.~F.} \bibnamefont{Heinz}}, \bibnamefont{and}
  \bibinfo{author}{\bibfnamefont{J.}~\bibnamefont{Shan}}, \bibinfo{journal}{Nat
  Mater} \textbf{\bibinfo{volume}{12}}, \bibinfo{pages}{207}
  (\bibinfo{year}{2013}).

\bibitem[{\citenamefont{{Yu} and {Wu}}(2014)}]{Yu:2014uq}
\bibinfo{author}{\bibfnamefont{T.}~\bibnamefont{{Yu}}} \bibnamefont{and}
  \bibinfo{author}{\bibfnamefont{M.~W.} \bibnamefont{{Wu}}},
  \bibinfo{journal}{ArXiv e-prints}  (\bibinfo{year}{2014}),
  \eprint{1401.0047}.

\bibitem[{\citenamefont{{Yu} et~al.}(2014)\citenamefont{{Yu}, {Liu}, {Gong},
  {Xu}, and {Yao}}}]{Yu:2014fk}
\bibinfo{author}{\bibfnamefont{H.}~\bibnamefont{{Yu}}},
  \bibinfo{author}{\bibfnamefont{G.}~\bibnamefont{{Liu}}},
  \bibinfo{author}{\bibfnamefont{P.}~\bibnamefont{{Gong}}},
  \bibinfo{author}{\bibfnamefont{X.}~\bibnamefont{{Xu}}}, \bibnamefont{and}
  \bibinfo{author}{\bibfnamefont{W.}~\bibnamefont{{Yao}}},
  \bibinfo{journal}{ArXiv e-prints}  (\bibinfo{year}{2014}),
  \eprint{1401.0667}.

\bibitem[{\citenamefont{Maialle et~al.}(1993)\citenamefont{Maialle,
  de~Andrada~e Silva, and Sham}}]{maialle93}
\bibinfo{author}{\bibfnamefont{M.}~\bibnamefont{Maialle}},
  \bibinfo{author}{\bibfnamefont{E.}~\bibnamefont{de~Andrada~e Silva}},
  \bibnamefont{and} \bibinfo{author}{\bibfnamefont{L.}~\bibnamefont{Sham}},
  \bibinfo{journal}{Phys. Rev. B} \textbf{\bibinfo{volume}{47}},
  \bibinfo{pages}{15776} (\bibinfo{year}{1993}).

\bibitem[{\citenamefont{Bir and Pikus}(1974)}]{birpikus_eng}
\bibinfo{author}{\bibfnamefont{G.~L.} \bibnamefont{Bir}} \bibnamefont{and}
  \bibinfo{author}{\bibfnamefont{G.~E.} \bibnamefont{Pikus}},
  \emph{\bibinfo{title}{Symmetry and Strain-induced Effects in Semiconductors}}
  (\bibinfo{publisher}{Wiley/Halsted Press}, \bibinfo{year}{1974}).

\bibitem[{\citenamefont{Denisov and Makarov}(1973)}]{denisovmakarov}
\bibinfo{author}{\bibfnamefont{M.~M.} \bibnamefont{Denisov}} \bibnamefont{and}
  \bibinfo{author}{\bibfnamefont{V.~P.} \bibnamefont{Makarov}},
  \bibinfo{journal}{Physica Status Solidi (b)} \textbf{\bibinfo{volume}{56}},
  \bibinfo{pages}{9} (\bibinfo{year}{1973}).

\bibitem[{\citenamefont{Koster et~al.}(1963)\citenamefont{Koster, Wheeler,
  Dimmock, and Statz}}]{koster63}
\bibinfo{author}{\bibfnamefont{G.~F.} \bibnamefont{Koster}},
  \bibinfo{author}{\bibfnamefont{R.~G.} \bibnamefont{Wheeler}},
  \bibinfo{author}{\bibfnamefont{J.~O.} \bibnamefont{Dimmock}},
  \bibnamefont{and} \bibinfo{author}{\bibfnamefont{H.}~\bibnamefont{Statz}},
  \emph{\bibinfo{title}{Properties of the thirty-two point groups}}
  (\bibinfo{publisher}{MIT Press}, \bibinfo{year}{1963}).

\bibitem[{\citenamefont{Korm\'anyos et~al.}(2013)\citenamefont{Korm\'anyos,
  Z\'olyomi, Drummond, Rakyta, Burkard, and Fal'ko}}]{Kormanyos:2013dq}
\bibinfo{author}{\bibfnamefont{A.}~\bibnamefont{Korm\'anyos}},
  \bibinfo{author}{\bibfnamefont{V.}~\bibnamefont{Z\'olyomi}},
  \bibinfo{author}{\bibfnamefont{N.~D.} \bibnamefont{Drummond}},
  \bibinfo{author}{\bibfnamefont{P.}~\bibnamefont{Rakyta}},
  \bibinfo{author}{\bibfnamefont{G.}~\bibnamefont{Burkard}}, \bibnamefont{and}
  \bibinfo{author}{\bibfnamefont{V.~I.} \bibnamefont{Fal'ko}},
  \bibinfo{journal}{Phys. Rev. B} \textbf{\bibinfo{volume}{88}},
  \bibinfo{pages}{045416} (\bibinfo{year}{2013}).

\bibitem[{\citenamefont{Liu et~al.}(2013)\citenamefont{Liu, Shan, Yao, Yao, and
  Xiao}}]{Liu:2013if}
\bibinfo{author}{\bibfnamefont{G.-B.} \bibnamefont{Liu}},
  \bibinfo{author}{\bibfnamefont{W.-Y.} \bibnamefont{Shan}},
  \bibinfo{author}{\bibfnamefont{Y.}~\bibnamefont{Yao}},
  \bibinfo{author}{\bibfnamefont{W.}~\bibnamefont{Yao}}, \bibnamefont{and}
  \bibinfo{author}{\bibfnamefont{D.}~\bibnamefont{Xiao}},
  \bibinfo{journal}{Phys. Rev. B} \textbf{\bibinfo{volume}{88}},
  \bibinfo{pages}{085433} (\bibinfo{year}{2013}).

\bibitem[{\citenamefont{Kosmider et~al.}(2013)\citenamefont{Kosmider,
  Gonz\'alez, and Fern\'andez-Rossier}}]{Kosmider:2013a}
\bibinfo{author}{\bibfnamefont{K.}~\bibnamefont{Kosmider}},
  \bibinfo{author}{\bibfnamefont{J.~W.} \bibnamefont{Gonz\'alez}},
  \bibnamefont{and}
  \bibinfo{author}{\bibfnamefont{J.}~\bibnamefont{Fern\'andez-Rossier}},
  \bibinfo{journal}{Phys. Rev. B} \textbf{\bibinfo{volume}{88}},
  \bibinfo{pages}{245436} (\bibinfo{year}{2013}).

\bibitem[{\citenamefont{Qiu et~al.}(2013)\citenamefont{Qiu, da~Jornada, and
  Louie}}]{Qiu:2013fe}
\bibinfo{author}{\bibfnamefont{D.~Y.} \bibnamefont{Qiu}},
  \bibinfo{author}{\bibfnamefont{F.~H.} \bibnamefont{da~Jornada}},
  \bibnamefont{and} \bibinfo{author}{\bibfnamefont{S.~G.} \bibnamefont{Louie}},
  \bibinfo{journal}{Phys. Rev. Lett.} \textbf{\bibinfo{volume}{111}},
  \bibinfo{pages}{216805} (\bibinfo{year}{2013}).

\bibitem[{\citenamefont{Agranovich and Ginzburg}(1984)}]{agr_ginz}
\bibinfo{author}{\bibfnamefont{V.}~\bibnamefont{Agranovich}} \bibnamefont{and}
  \bibinfo{author}{\bibfnamefont{V.}~\bibnamefont{Ginzburg}},
  \emph{\bibinfo{title}{Crystal optics with spatial dispersion, and excitons}}
  (\bibinfo{publisher}{Springer-Verlag (Berlin and New York)},
  \bibinfo{year}{1984}).

\bibitem[{\citenamefont{Ivchenko}(2005)}]{ivchenko05a}
\bibinfo{author}{\bibfnamefont{E.~L.} \bibnamefont{Ivchenko}},
  \emph{\bibinfo{title}{Optical spectroscopy of semiconductor nanostructures}}
  (\bibinfo{publisher}{Alpha Science, Harrow UK}, \bibinfo{year}{2005}).

\bibitem[{\citenamefont{Goupalov et~al.}(1998)\citenamefont{Goupalov, Ivchenko,
  and Kavokin}}]{goupalov98}
\bibinfo{author}{\bibfnamefont{S.~V.} \bibnamefont{Goupalov}},
  \bibinfo{author}{\bibfnamefont{E.~L.} \bibnamefont{Ivchenko}},
  \bibnamefont{and} \bibinfo{author}{\bibfnamefont{A.~V.}
  \bibnamefont{Kavokin}}, \bibinfo{journal}{JETP}
  \textbf{\bibinfo{volume}{86}}, \bibinfo{pages}{388} (\bibinfo{year}{1998}).

\bibitem[{\citenamefont{Goupalov et~al.}(2003)\citenamefont{Goupalov,
  Lavallard, Lamouche, and Citrin}}]{goupalov:electrodyn}
\bibinfo{author}{\bibfnamefont{S.~V.} \bibnamefont{Goupalov}},
  \bibinfo{author}{\bibfnamefont{P.}~\bibnamefont{Lavallard}},
  \bibinfo{author}{\bibfnamefont{G.}~\bibnamefont{Lamouche}}, \bibnamefont{and}
  \bibinfo{author}{\bibfnamefont{D.~S.} \bibnamefont{Citrin}},
  \bibinfo{journal}{{Physics of the Solid State}}
  \textbf{\bibinfo{volume}{45}}, \bibinfo{pages}{768} (\bibinfo{year}{2003}).

\bibitem[{\citenamefont{Kiselev and Zhilich}(1972)}]{zhilich72:eng}
\bibinfo{author}{\bibfnamefont{V.~A.} \bibnamefont{Kiselev}} \bibnamefont{and}
  \bibinfo{author}{\bibfnamefont{A.~G.} \bibnamefont{Zhilich}},
  \bibinfo{journal}{Sov. Phys. Solid State} \textbf{\bibinfo{volume}{13}},
  \bibinfo{pages}{2008} (\bibinfo{year}{1972}).

\bibitem[{\citenamefont{Kiselev and Zhilich}(1974)}]{zhilich74:eng}
\bibinfo{author}{\bibfnamefont{V.~A.} \bibnamefont{Kiselev}} \bibnamefont{and}
  \bibinfo{author}{\bibfnamefont{A.~G.} \bibnamefont{Zhilich}},
  \bibinfo{journal}{Sov. Phys. - Semicond.} \textbf{\bibinfo{volume}{8}}
  (\bibinfo{year}{1974}).

\bibitem[{\citenamefont{Dyakonov and Perel'}(1972)}]{dyakonov72}
\bibinfo{author}{\bibfnamefont{M.}~\bibnamefont{Dyakonov}} \bibnamefont{and}
  \bibinfo{author}{\bibfnamefont{V.}~\bibnamefont{Perel'}},
  \bibinfo{journal}{Sov. Phys. Solid State} \textbf{\bibinfo{volume}{13}},
  \bibinfo{pages}{3023} (\bibinfo{year}{1972}).

\bibitem[{\citenamefont{Meier and Zakharchenya}(1984)}]{opt_or_book}
\bibinfo{editor}{\bibfnamefont{F.}~\bibnamefont{Meier}} \bibnamefont{and}
  \bibinfo{editor}{\bibfnamefont{B.}~\bibnamefont{Zakharchenya}}, eds.,
  \emph{\bibinfo{title}{Optical orientation}}
  (\bibinfo{publisher}{Horth-Holland, Amsterdam}, \bibinfo{year}{1984}).

\bibitem[{\citenamefont{Korn et~al.}(2011)\citenamefont{Korn, Heydrich, Hirmer,
  Schmutzler, and Sch{\"u}ller}}]{KornMoS2}
\bibinfo{author}{\bibfnamefont{T.}~\bibnamefont{Korn}},
  \bibinfo{author}{\bibfnamefont{S.}~\bibnamefont{Heydrich}},
  \bibinfo{author}{\bibfnamefont{M.}~\bibnamefont{Hirmer}},
  \bibinfo{author}{\bibfnamefont{J.}~\bibnamefont{Schmutzler}},
  \bibnamefont{and}
  \bibinfo{author}{\bibfnamefont{C.}~\bibnamefont{Sch{\"u}ller}},
  \bibinfo{journal}{Applied Physics Letters} \textbf{\bibinfo{volume}{99}},
  \bibinfo{eid}{102109} (\bibinfo{year}{2011}).

\bibitem[{\citenamefont{Jones et~al.}(2013)\citenamefont{Jones, Yu, Ghimire,
  Wu, Aivazian, Ross, Zhao, Yan, Mandrus, Xiao et~al.}}]{Jones:2013tg}
\bibinfo{author}{\bibfnamefont{A.~M.} \bibnamefont{Jones}},
  \bibinfo{author}{\bibfnamefont{H.}~\bibnamefont{Yu}},
  \bibinfo{author}{\bibfnamefont{N.~J.} \bibnamefont{Ghimire}},
  \bibinfo{author}{\bibfnamefont{S.}~\bibnamefont{Wu}},
  \bibinfo{author}{\bibfnamefont{G.}~\bibnamefont{Aivazian}},
  \bibinfo{author}{\bibfnamefont{J.~S.} \bibnamefont{Ross}},
  \bibinfo{author}{\bibfnamefont{B.}~\bibnamefont{Zhao}},
  \bibinfo{author}{\bibfnamefont{J.}~\bibnamefont{Yan}},
  \bibinfo{author}{\bibfnamefont{D.~G.} \bibnamefont{Mandrus}},
  \bibinfo{author}{\bibfnamefont{D.}~\bibnamefont{Xiao}}, \bibnamefont{et~al.},
  \bibinfo{journal}{Nat Nano} \textbf{\bibinfo{volume}{8}},
  \bibinfo{pages}{634} (\bibinfo{year}{2013}).

\end{thebibliography}

\end{document}